# Control of a hair bundle's mechanosensory function by its mechanical load

Joshua D. Salvi*, Dáibhid Ó Maoiléidigh*, Brian A. Fabella, Melanie Tobin[†], and A. J. Hudspeth

Howard Hughes Medical Institute and Laboratory of Sensory Neuroscience, The Rockefeller University, 1230 York Avenue, New York, New York, 10065, USA

[*] These authors contributed equally to the work.

[†] Current address:   Laboratoire Physico-Chimie Curie, Centre National de la Recherche Scientifique, Institut Curie, 26 rue d'Ulm, F-75248 Cedex 05, Paris, France

**Abstract**

Hair cells, the sensory receptors of the internal ear, subserve different functions in various receptor organs: they detect oscillatory stimuli in the auditory system but transduce constant and step stimuli in the vestibular and lateral line systems. We show that a hair cell's function can be controlled experimentally by adjusting its mechanical load. By making bundles from a single organ operate as any of four distinct types of signal detector, we demonstrate that altering only a few key parameters can fundamentally change a sensory cell's role. The motions of a single hair bundle can resemble those of a bundle from the amphibian vestibular system, the reptilian auditory system, or the mammalian auditory system, demonstrating an essential similarity of bundles across species and receptor organs.

**Significance statement**

Hair bundles are the sensory antennae that detect different types of mechanical signals in diverse sensory systems across organisms. Here we design and employ a mechanical-load clamp to show that the mechanical properties of hair bundles and their accessory structures dictate their sensory behaviors. This study reveals both the versatility and an essential similarity of hair bundles across sensory organs by demonstrating how the same organelle can be used to detect a wide gamut of signals. These observations of a particular sensory structure reveal a general principle that may be utilized by both biological and artificial systems: a nonlinear system can be controlled to serve many different functions by adjusting only a few key parameters.

# Introduction

The hair bundles of vertebrates are mechanosensory organelles responsible for detecting sounds in the auditory system, linear and angular accelerations in the vestibular system, and water movements and pressure gradients in the lateral-line system of fishes and amphibians *(1,2)*. In each instance an appropriate stimulus deflects the bundles, depolarizing the hair cells from which they emerge *(3)*. Hair bundles are not simply passive detectors, however, for they actively amplify their responses to mechanical stimulation *(4,5)*. Hair-bundle motility contributes to an active process that endows the auditory system with the ability to detect stimuli with energies near that of thermal fluctuations (*6*), to distinguish tones that differ by less than 0.2% in frequency *(7)*, and to accommodate inputs varying over a millionfold range in amplitude *(4,8-11)*.

Auditory, vestibular, and lateral-line organs respond to distinct patterns of mechanical input. The mechanical properties and environments of hair bundles differ correspondingly between different organisms, among receptor organs, and with the tonotopic position along individual auditory organs *(12-15)*. A mathematical model predicts that the response of a hair bundle is regulated both by its intrinsic mechanical characteristics and by its mechanical load *(16)*. Motivated by this proposition, we investigated how the load stiffness, or force per unit displacement, and constant force imposed on a bundle controls its dynamics and response to external perturbations.

# Results

**Mapping the hair bundle's experimental state diagram**

The hair bundle's state diagram characterizes its behavior for different combinations of two control parameters: load stiffness and constant force. These control parameters describe the mechanical load imposed on a hair bundle within a sensory organ. A theoretical model of hair bundle dynamics predicts the qualitative structure of the state diagram (Fig. 1A-C). To test this prediction experimentally, we varied the mechanical load imposed on an individual hair bundle and monitored its displacement. The load was delivered to an individual hair bundle by attaching the tip of a flexible glass fiber to the bundle's top and by employing a piezoelectric actuator to displace the fiber's base. To control the load, we used a real-time processor that compared the bundle's actual position, measured by a photomicrometer system, with that specified by a command signal, and then provided feedback to the actuator to minimize the difference between the two (Fig. 1D). By adjusting the strength of the system's feedback and the commanded position, we independently manipulated the load stiffness and the constant force, which allowed us to vary the bundle's operating point to reveal its experimental state diagram (Fig. 1, Supplementary Fig. 1, and the *Mechanical-load clamp* section in *Materials and Methods*).

At each operating point we classified a hair bundle's motion as oscillatory or quiescent based on the distribution of its displacement (*State-diagram mapping and analysis* section in *Materials and Methods*). In accord with previous manipulations of applied force *(17)*, we found that hair bundles oscillated spontaneously for operating points in certain regions of the experimental state diagram but were quiescent for others. The activity and mechanical properties of hair bundles depended on their diameter, which resulted in oscillatory regions of varying extent (Fig. 2 and Supplementary Fig. 2). For the lowest values of load stiffness, small bundles exhibited multimodal oscillations (Fig. 2A) *(18)*. The limited range of operating points for which the clamp was stable permitted exploration of only the region of spontaneous oscillation for the largest hair bundles (Fig. 2I-L and Supplementary Figures 2G-C and 3C).

In agreement with simulations, an increase in the stiffness was correlated with a rise in the frequency of oscillation for many bundles and with a decrease in the amplitude of oscillation in all cases (Fig. 2, Supplementary Table 1, and Supplementary Figures 2, 3 and 4) *(16)*. The decrease in the amplitude of oscillation for a given bundle persisted for the duration of an experiment (Supplementary Fig. 3A-B). The region of spontaneous oscillation was abolished upon addition of gentamicin, a drug that blocks mechanoelectrical-transduction channels (Supplementary Fig. 3C-F). This result implies that the architecture of the experimental state diagram reflects active hair-bundle motility, which depends upon functional channels.

**Response to sinusoidal stimulation**

The principal role of hair cells in the auditory system is the detection of periodic forces derived from sound. Although modeling suggests that a quiescent hair bundle should respond with maximal frequency tuning and nonlinearity at the boundary of the region of spontaneous oscillation *(11,19-22)*, no experimental study has systematically examined bundles near this critical locus. We therefore inquired how hair bundles situated at various positions in the experimental state diagram respond to sinusoidal stimulation over a range of amplitudes and frequencies.

The boundary between the regions of spontaneous activity and quiescence lay between operating points for which the bundle's behavior could be clearly classified as either oscillatory or quiescent based on the distribution of its displacement. When poised near this border, a hair bundle exhibited resonant frequency tuning: its phase-locked response—the magnitude of its average oscillation amplitude at the stimulus frequency—in response to sinusoidal stimulation displayed a clear peak (Fig. 3A). As the bundle's operating point was displaced into the quiescent region, the response at this resonant frequency progressively diminished. The sharpness of tuning, quantified by the quality factor of the resonance, was greatest on the oscillatory side of the boundary and in one instance approached that of auditory organs *(10,30)* (Fig. 3A and Supplementary Fig. 5A,C).

When we exposed a hair cell's apical surface to gentamicin to abolish active hair-bundle motility, a hair bundle's tuning diminished at all operating points; the response at the original resonant frequency collapsed to the noise floor (Fig. 3A and Supplementary Fig. 5A,C). This result confirmed that the peak response and sharpness of tuning stem from active hair-bundle motility that augments the bundle's mechanical response. Indeed, the response of an untreated hair bundle exceeded that of a treated bundle for most stiffness values and stimulus frequencies.

Unlike the response of an untreated bundle, that of a treated bundle was insensitive to the load stiffness and resembled that of a low-pass filter.

The phase of an active hair bundle's response led that of the stimulus for frequencies below the bundle's resonant frequency and lagged for frequencies exceeding that value (Fig. 3B and Supplementary Fig. 5B,D) *(11)*. The phase lead diminished far from and on the quiescent side of the border of spontaneous oscillation and disappeared upon application of gentamicin, indicating that the lead was generated by active hair-bundle motility (Fig. 3B).

A hair bundle's sensitivity—its phase-locked response divided by the magnitude of the stimulus force—decreases for forces of increasing amplitude (Fig. 3E-F and Supplementary Fig. 6) *(11)*. This change in the sensitivity is greatest for spontaneously oscillating bundles and the sensitivity appears to become independent of the operating point for large forcing. The sensitivity is largest for operating points for which the hair bundle oscillates, although the bundle is most strongly entrained when it operates on the quiescent side of and near the border of spontaneous oscillation for forces of intermediate magnitude. Unlike previous experimental and theoretical work that described a linear response for large-amplitude forcing (48), no such effect was found in the present work owing to an experimental limit on the maximal stimulus force possible.

To compare the experimental behavior of hair bundles with the predictions of a dynamical model of hair-bundle activity *(16)*, we conducted simulations of the responses expected in the presence of thermal noise. The model's response agreed qualitatively with the experimental observations (Fig. 3C-D,G-H). Near the boundary of the oscillatory region, the entrained responses accorded with those predicted for a system operating at the edge of an oscillatory instability *(22-25)*, for which the sensitivity decreases as the magnitude of the force to the power of -2/3.

Far from the border of the oscillatory region, a hair bundle's maximal response to weak stimulation was smaller and the response as a function of frequency was less sharply tuned than near the border (Fig. 4A). Hypothesizing that this observation reflected reduced entrainment to the stimulus, we assessed how well the responses followed the stimuli by computing the vector strength at the resonant frequency for a range of load stiffnesses. The vector strength peaked at the edge and on the quiescent side of the boundary of spontaneous oscillations (Fig. 4B). To better characterize entrainment, we delivered stimuli of both increasing force and increasing frequency to oscillatory hair bundles subjected to three load stiffnesses. Entrainment to small stimulus forces was maximal for operating points nearest the boundary of oscillation (Fig. 4C). This dependence of vector strength on the operating point diminished for stimulation away from the bundle's characteristic frequency (Fig. 4C-E). In agreement with published observations *(26,27)*, entrainment was frequency-dependent for a range of stimulus forces. A map of vector strength computed across a range of stimulus forces and frequencies disclosed a phase-locked region that depended on load stiffness. Although experiments were limited by the bundle's low resonant frequency, the entrainment region appeared to become more sharply tuned as the load stiffness increased (Fig 4F-H), an effect that was previously obscured by the use of different hair bundles for each load stiffness *(26)*. The responses to periodic forcing together indicate that a hair bundle's capacity to encode auditory stimuli depends upon its mechanical load.

**Response to pulsatile stimulation**

Hair bundles are often investigated *in vitro* with pulses of displacement or force *(1,8,28)* that additionally correspond to the stimulation of some vestibular organs *in vivo (29)*. Because we expected a bundle's response to such stimuli to depend upon its load *(16)*, we recorded responses to force pulses from hair bundles subjected to mechanical loads representing operating points within and outside their regions of spontaneous oscillation.

As the constant force applied to a bundle became more negative, hair-bundle oscillations decreased in frequency but not in amplitude until they ceased altogether (Fig. 5A), in agreement with the results of previous studies that employed a low load stiffness *(27)*. Positive force pulses delivered to a hair bundle subjected to successively more negative constant forces first induced oscillations and subsequently evoked twitches *(1,39)*, rapid movements that transiently exerted negative forces on the stimulus fiber (Fig. 5A). A second bundle held under a negative constant force responded to a positive pulse by moving less than the stimulus fiber's base (Fig. 5B). When the bundle was subjected to a still more negative constant force, however, the bundle's displacement in response to a pulse exceeded that of the stimulus fiber's base. During this movement, the force delivered to the hair bundle was negative in sign, implying that the hair bundle exerted a positive force on the fiber (Fig. 5C). This phenomenon, which had previously been observed only once before for an outer hair cell's bundle in the rat's cochlea *(28)*, demonstrates that amphibian vestibular hair bundles can be induced to respond qualitatively like mammalian auditory bundles despite their different morphologies. The behavior arises from the nonlinear stiffness of a bundle owing to channel gating, a feature conserved in bundles across receptor organs.

## Discussion

From the level of molecules to that of ecosystems, a cardinal feature of life is the detection of environmental perturbations by sensory systems. Owing to the complexity of cells and organisms, however, mathematical descriptions that make testable predictions about such systems are rare. Moreover, experimental tools are seldom available to test theoretical predictions about biological dynamics. The present work represents an exception to these generalizations. On the basis of a simple dynamical model of hair-bundle activity, we have predicted the effects of mechanical loading on bundles *(16)*. The present experimental results accord with these predictions. For example, the experimental state diagram revealed by mechanical-load clamping displays a bounded locus of spontaneous activity. Within that area, the frequency of oscillation increases and the amplitude decreases as the boundary is approached by increasing the load stiffness. The changes in the hair bundle's response to sinusoidal and pulsatile stimuli as its operating point varies are also in qualitative agreement with the model's predictions.

We observed that increasing the stiffness and constant force confronting a hair bundle drives it from the region of spontaneous oscillation into a domain of quiescence. The boundary between the two regimes represents a bifurcation, that is, a dramatic change in behavior in response to continuous variation of one or more control parameters. Several observations indicate that the boundary associated with large values of the load stiffness constitutes a line of supercritical Hopf bifurcations *(11,16,20,26,38)*. Both the tuning and degree of entrainment of a bundle are maximal near the bifurcation; the frequency of spontaneous oscillations on one side matches the resonant frequency on the quiescent side; and the amplitude of spontaneous movement grows continuously from zero as the operating point progresses into the region of spontaneous oscillation. For low values of the load stiffness, however, the amplitude of spontaneous oscillation does not change as the bifurcation is approached, a behavior more consistent with a fold of limit cycles or an infinite-period bifurcation *(32,33)*. The model for hair-bundle mechanics predicts this change in the type of bifurcation as the stiffness is varied *(16)*.

The characteristic frequencies measured near the supercritical bifurcation accord with those of saccular afferent axons *in vivo (34,35)*, suggesting that hair bundles normally reside at operating points near the high-stiffness arc of bifurcations. Consistent with this idea, the elastic load imposed experimentally at the bifurcation resembled the stiffness of the otolithic membrane that ordinarily confronts a saccular hair bundle *(36)*. This result reinforces the thesis that the behavior of hair bundles *in vivo* is dictated by their mechanical loads. We also predict that reduction of these elastic loads *in vivo* will affect the oscillatory behavior of hair bundles, for example by increasing the prevalence and amplitude of otoacoustic emissions.

Although the sharpness of tuning that we measured was less than that of the mammalian cochlea *(37)*, it largely accounts for frequency tuning in the amphibian auditory system *(30)*. Moreover, the sharpness of tuning and the degree of nonlinearity found here represent lower estimates owing to the limited frequency resolution of the recordings. In agreement with theory

*(16)*, hair bundles situated at operating points distant from the oscillatory region lost their resonant character and behaved as low-pass filters.

Within a receptor organ, a hair bundle must counter viscous damping in order to minimize the loss of stimulus energy. Our results indicate that the bundle accomplishes this task when poised near its boundary of spontaneous activity. Although all bundles may potentially exhibit this behavior, mechanical loading by accessory structures controls their ability to amplify external signals. It remains to be determined how a hair bundle *in vivo* finds operating conditions that foster optimal responsiveness *(23)*.

Our results demonstrate an essential similarity of hair bundles, whose responsiveness in various receptor organs is controlled by mechanical loading. Although previous studies used different stimulus fibers to investigate the effects of stiffness and constant force on different hair bundles *(26,38)*, our approach revealed multiple mechanosensory modes in individual bundles. Depending upon its operating point, an individual bundle may twitch and oscillate like those in amphibian and reptilian receptor organs *(39)* or overshoot the stimulus like those in the mammalian cochlea *(28)*. Although hair bundles in the mammalian cochlea detect frequencies extending two or three orders of magnitude greater than those detected by the bullfrog's sacculus, these results indicate that bundles from both organs rely on the same essential mechanisms. Adjustments to these mechanisms—the rate of adaptation and the degree of nonlinearity—regulate the speed and range of responsiveness. It is probable that the physical properties of hair bundles and their accessory structures have evolved to adjust the functions of different receptor organs within a range of organisms.

In contrast to a manufactured device that is designed to produce or respond to signals in a stereotypical fashion, a single hair bundle can function in various capacities. Here we show that an individual hair bundle can behave like any of four different devices. A bundle may generate spontaneous oscillations like an oscillator that produces repetitive square or sine waves. It can resonate with high frequency resolution like a resonant circuit that responds to one frequency with greater amplitude than to any other. By attenuating high-frequency stimuli, a bundle may serve as a low-pass filter that attenuates signals above a cutoff frequency. Finally, by twitching at the onset of a pulse displacement, a bundle can mimic a step detector that identifies discontinuities in incoming signals. These observations for a particular sensory organelle reveal a general principle that may be utilized by both biological and artificial systems: a nonlinear system can be controlled to serve many different functions by adjusting only a few key parameters.

## Materials and Methods

**Experimental preparation.** All procedures were approved by the Institutional Animal Care and Use Committee of The Rockefeller University. Experiments were performed at 21 ˚C on hair cells from the saccular maculae of adult bullfrogs, *Rana catesbeiana*. Each dissected sacculus was placed in oxygenated artificial perilymph containing 114 mM $Na^+$, 2 mM $K^+$, 2 mM $Ca^{2+}$, 118 mM $Cl^-$, 5 mM HEPES, and 3 mM D-glucose. After isolation from the labyrinth and removal of otoconia, the saccular macula was sealed over a 1 mm hole in a 12 mm disk of aluminum foil with *n*-butyl cyanoacrylate (Vetbond, 3M, St. Paul, MN) to form a partition in a two-compartment chamber. The apical surface was exposed to 67 mg·$l^{-1}$ of protease (type XXIV, Sigma, St. Louis, MO) for 35 min at 21 ˚C to loosen the otolithic membrane, which was then removed with an eyelash. During recordings the lower chamber contained oxygenated artificial perilymph and the upper chamber held oxygenated artificial endolymph containing 2 mM $Na^+$, 118 mM $K^+$, 250 μM $Ca^{2+}$, 118 mM $Cl^-$, 5 mM HEPES, and 3 mM D-glucose. Both solutions had a pH of 7.3 and an osmotic strength of 230 mOsmol·$kg^{-1}$.

**Microscopic apparatus.** Hair bundles were visualized by differential-interference-contrast optics through a 60X water-immersion objective lens of numerical aperture 0.9 on an upright microscope (BX51WI, Olympus, Tokyo, Japan). To detect spontaneously active hair bundles, we directed the image through a 0.35X or 4.0X telescope to a charge-coupled-device camera and a video processor (Argus-20, Hamamatsu Photonics K. K., Hamamatsu City, Japan). Digital subtraction of each frame from the average of the previous one to five frames eased the detection of hair-bundle oscillations. An infrared-reflecting mirror (21002b, Chroma Technology, Bellows Falls, VT) and a broadband interference filter (585 ± 30 nm; #220494, Chroma Technology) protected the tissue from photodamage. For experimental measurements, the polarizer and filter were removed from the light path and the sample was illuminated at 630 nm with a 900 mW light-emitting diode (UHP-Mic-LED-630, Prizmatix, Givat-Shmuel, Israel).

**Mechanical stimulation.** Mechanical stimuli were delivered by flexible glass fibers fabricated from borosilicate capillaries 1.2 mm in external diameter (TW120-3, World Precision Instruments, Sarasota, FL). After a capillary had been narrowed by an electrode puller (P-2000, Sutter Instruments, Novato, CA), a 120 V solenoid pulling at a right angle created a solid fiber no more than 100 μm in length and 0.5-0.8 μm in diameter. To enhance its optical contrast, each fiber was sputter-coated with gold-palladium (Hummer 6.2, Anatech, Hayward, CA). To improve its attachment to the kinociliary bulb, each fiber was treated for 15 min with 200 μg·l$^{-1}$ concanavalin A (type IV, Sigma, St. Louis, MO).

To determine a fiber's stiffness and drag coefficient, we analyzed its thermal fluctuations while submerged in water. We fit the power spectrum $S_X$ as a function of frequency $f$ from a 30 s record to the Lorentzian relation (49)

$$S_X(f) = \frac{a}{f_0^2 + f^2};  \qquad (1)$$

$$\xi_{SF} = \frac{k_B T}{\pi^2 a};  \qquad (2)$$

$$K_{SF} = 2\pi \xi_{SF} f_0. \qquad (3)$$

Here $a$ is a fitting parameter, $f_0$ is the half-power frequency, $k_B$ is Boltzmann's constant, $T$ is the temperature, $\xi_{SF}$ is the fiber's drag coefficient, and $K_{SF}$ is the fiber's stiffness. Fibers had stiffnesses of 50-600 μN·m$^{-1}$ and drag coefficients of 25-80 nN·s·m$^{-1}$.

The base of each stimulus fiber was secured to a high-frequency piezoelectric actuator (PA 4/12, Piezosystem Jena GmbH, Jena, Germany) driven by an 800 mA amplifier (ENV 800,

Piezosystem Jena). The actuator was mounted on a micromanipulator (ROE-200, Sutter Instruments) for positioning of the fiber's tip.

The tip of a horizontally mounted stimulus fiber was tightly coupled to the kinociliary bulb of an individual hair bundle. Each hair bundle was classified by its diameter at the insertion into the cuticular plate. A small bundle was estimated to have a diameter of less than 2 µm and about 20 stereocilia; a medium bundle was 2-4 µm in diameter and encompassed approximately 40 stereocilia, whereas a large bundle of more than 4 µm contained around 60 stereocilia.

**Photometric recording.** The motion of a hair bundle was tracked by imaging the stimulus fiber's tip on a dual photodiode at a magnification of 1,350X. The output of the photodiode was then relayed through a low-pass filter with a cutoff frequency of 2 kHz (BM8, Kemo Limited, Dartford, United Kingdom). The sensitivity of the photodiode system was calibrated by independently translating the fiber's image in 20 µm steps with a mirror coupled to a second piezoelectric actuator driven by a 300 mA amplifier (PA 120/14 SG and ENV 300 SG, Piezosystem Jena). This actuator was calibrated by a heterodyne interferometer (OFV 501, Polytec GmbH, Waldbronn, Germany).

**Mechanical-load clamp.** In most clamp systems, such as the venerable voltage clamp, a negative feedback loop holds one experimental variable fixed while the conjugate variable is measured. Under displacement-clamp conditions, for example, a hair bundle is maintained at a commanded position while the ensuing force is evaluated *(42,43)*. A force clamp inverts this relationship: feedback imposes a constant force while the displacement is determined.

In the present experiments we implemented a generalization of this procedure, load-clamping, in which the feedback system simultaneously imposes on a hair bundle two conditions that mimic the bundle's environment *in vivo*. The system serves as a stiffness clamp that imposes a specified elastic load and at the same time acts as a force clamp that applies a commanded constant, sinusoidal, or pulsatile force. Load-clamping is possible because contemporary computers can solve the necessary differential equations on a timescale shorter than the mechanical relaxation time of the stimulus fiber and attached hair bundle *(40,44)*. In a further generalization of the approach, we intend to examine the hypothesized effects of inertia and drag on hair bundles, two parameters that may for example play a greater role in vestibular and auditory organs *(16)*. An extension of the present system shall include additional mechanical control parameters and is currently under development.

At the point of contact between a stimulus fiber and hair bundle, the elastic force produced by the flexion of the fiber is balanced by the sum of the elastic and drag forces associated with the fiber and bundle:

$$K_{SF}(\Delta - X) = \xi_{XX}\dot{X} + \xi_{\Delta X}\dot{\Delta} - \xi_{HB}\dot{X} + K_{HB}X + F_A. \quad (4)$$

Here $K_{SF}$ is the stiffness of the stimulus fiber and $K_{HB}$ is that of the hair bundle, $\Delta$ is the position of the fiber's base and $X$ is that of the hair bundle, $\xi_{XX}$ is the drag coefficient of the stimulus fiber owing to motion at the fiber's tip and $\xi_{\Delta X}$ is that owing to motion at the fiber's base, and $\xi_{HB}$ is the drag coefficient of the hair bundle. $F_A$ represents any active or nonlinear force produced by the hair bundle. Note that all inertial effects of the bundle and fiber are assumed to be small.

The photodiode's output voltage $V_D$ is linearly related to the hair bundle's position by a coefficient $\alpha$: $V_D = \alpha X$. If the clamp accomplishes a commanded displacement $X_C$, the photodiode's output voltage is $V_C = \alpha X_C$. The error signal at the clamp's amplifier is therefore

$$V_E = V_C - V_D = \alpha(X_C - X). \quad (5)$$

This signal is amplified by the gain $G$ to generate an output signal $V_O$ that is delivered to the piezoelectric stimulator,

$$V_O = GV_E = \alpha G(X_C - X). \quad (6)$$

The stimulator's displacement output is linearly related to its input signal by a coefficient $\beta$, $\Delta = \beta V_O$, so the resultant displacement of the stimulus fiber's base is

$$\Delta = \alpha\beta G(X_C - X). \quad (7)$$

Combining Equations 4 and 7 yields

$$(\xi_{XX} - \alpha\beta G \xi_{\Delta X} + \xi_{HB})\dot{X} + [(1 + \alpha\beta G)K_{SF} + K_{HB}]X + F_A = \alpha\beta G(-\xi_{\Delta X}\dot{X}_C + K_{SF}X_C). \quad (8)$$

When $|\alpha\beta G \xi_{\Delta X}| \ll \xi_{XX} + \xi_{HB}$ and $|K_{SF}X_C| \gg |\xi_{\Delta X}\dot{X}_C|$, Equation 8 becomes

$$(\xi_{XX} + \xi_{HB})\dot{X} + (K_{EFF} + K_{HB})X + F_A = \alpha\beta G K_{SF} X_C, \qquad (9)$$

in which $K_{EFF} = (1 + \alpha\beta G)K_{SF}$ is the effective load stiffness due to the clamp. The condition $|\alpha\beta G\xi_{\Delta X}| \ll \xi_{XX} + \xi_{HB}$ is met for sufficiently small values of $\xi_{\Delta X}$ and $G$, in which $\alpha\beta \sim 1$ and $\xi_{HB} \geq 100$ nN·s·m$^{-1}$. The experiments presented here satisfy this condition, with $G$ typically less than one and never exceeding two. For the fibers used here *(45)*, $\xi_{XX}$ = 50-70 nN·s·m$^{-1}$ and $\xi_{\Delta X}$ = 30-40 nN·s·m$^{-1}$. When $K_{SF}$ = 100-350 µN·m$^{-1}$, the condition $|K_{SF} X_C| \gg |\xi_{\Delta X} \dot{X}_C|$ is satisfied for timescales greater than $\xi_{\Delta X}/K_{SF} <$ 0.1-0.3 ms. Additional forces owing to the drag from the base of the fiber are not significant for times exceeding this bound.

The displacement command may be used to apply various types of stimuli to the bundle. We may choose $X_C = X_{CC} + X_S \sin(\omega_s t) + X_P(t)$, such that $F_C = \alpha\beta G K_{SF} X_{CC}$ is a constant force, $F_S \sin(\omega_s t) = \alpha\beta G K_{SF} X_S \sin(\omega_s t)$ is a sinusoidal force of angular frequency $\omega_S$, and $F_P(t) = \alpha\beta G K_{SF} X_P(t)$ is a force pulse. Equation 9 then yields

$$(\xi_{XX} + \xi_{HB})\dot{X} + (K_{EFF} + K_{HB})X + F_A = F_C + F_S \sin(\omega_s t) + F_P(t). \qquad (10)$$

The clamp thus allows us to control the stiffness and various forms of stimulus force independently through adjustment of the proportional gain $G$ and command displacement $X_C$.

After poising a hair bundle at a particular operating point defined by the stiffness and offset force, we sometimes delivered sinusoidal force stimuli and recorded the bundle's response. Prior to each set of stimuli, the hair bundle was held at its operating point for 3 s to allow it to reach a steady state; stimuli were subsequently delivered for 15-50 cycles, depending upon the frequencies chosen. In other experiments we used a similar protocol to deliver 1 s force pulses.

**Clamp verification.** To demonstrate robust independent control of the stiffness and constant force with a mechanical-load clamp, we applied stimuli to a vertically mounted glass fiber that served as a simulacrum of a hair bundle. This arrangement provided a system with a known, linear stiffness for calibration and controls. For calibration purposes, steps were delivered for a series of forces and stiffnesses. For each constant force, the steady-state position of the fibers' tip is given by

$$X = \frac{\alpha\beta G X_C K_{SF}}{K_{SF}(1+\alpha\beta G)+K_{HB}} = \frac{F_C}{K_{EFF}+K_{HB}} \quad . \tag{11}$$

The validity of this relation is confirmed in Supplementary Fig. 1.

**Signal production and acquisition.** Stimuli were generated and data recorded by a host computer running programs written in LabVIEW (version 10.0, National Instruments, Austin, TX) with a sampling interval of 100 μs. For mechanical load-clamp experiments, signals were relayed to a target computer running the LabVIEW Real-Time operating system (version 10.0, National Instruments). To set a defined stiffness and constant force, the host computer adjusted the proportional gain and command displacement and transmitted those values to the target computer. The target computer then rapidly executed a short program that implemented Equation 6 and provided an appropriate signal to the piezoelectric stimulator.

**Gentamicin controls.** We evaluated the sensitivity of hair bundles to applied forces when oscillations were arrested. Transduction channels were blocked with 500 μM gentamicin sulfate applied in the upper chamber, a treatment whose reversibility allowed bundles to be reassessed after washout.

**State-diagram mapping and analysis.** To construct an experimental state diagram, we subjected a hair bundle to a set of load stiffnesses and constant forces, recording at each operating point for 2-8 s. The control parameters were specified for each operating point by a constant command displacement $X_C$ and proportional gain $G$. The duration was limited by the stability of the load clamp, which took about 3 min to recalibrate after each 1 min of recording, and by the time of 30 min during which the bundle's dynamics was unchanged by cellular deterioration. Here we describe the procedure for classifying and analyzing the operating points that constituted the bundle's experimental state diagram. The parameter values used for each diagram are listed in Supplementary Table 1; the analysis was performed using MATLAB (R2014a, 8.3.0.532).

If a hair bundle was oscillating at a particular operating point, then the distribution of its displacements displayed more than one peak. To analyze this distribution, we first removed slow drift in the time trace of bundle displacement by subtracting the time trace smoothed by moving averages over a time window of a fixed length. We then employed up to three statistical tests to determine whether the displacement distribution was multimodal. Hartigans' dip statistic is larger for multimodal distributions than for unimodal ones *(46)*. Using as the null distribution a normal distribution with the same mean and variance as the displacement distribution to make the test more sensitive, we defined a multimodality score to be equal to the dip statistic. Displacement distributions that differed statistically from normal, possessing a multimodality score exceeding a threshold, corresponded to oscillatory operating points. This procedure did not identify all the operating points at which a bundle oscillated, however, for noise could obscure the dips between peaks in a distribution, resulting in a distribution that was asymmetric or broad.

To determine if a distribution is excessively asymmetric we created two additional distributions. The right distribution was constructed by reflecting about the mean each displacement greater than the mean displacement; the left distribution was found in an analogous manner. This process yielded two symmetric distributions that corresponded to the mirroring of the left and right halves of the original distribution. The original distribution was considered asymmetric if the left distribution and right distribution were statistically distinct. We defined the asymmetry score as the Kolmogorov-Smirnov test statistic resulting from a comparison of the left and right distributions. A distribution was judged to be asymmetric if the score exceeded a threshold and was statistically significant.

Broad distributions were identified as those with negative excess kurtosis. We defined a thinness score as the excess kurtosis divided by its standard error. This score has a normal distribution for large samples *(47)*. A distribution was considered broad if the thinness score lay below a threshold and was statistically smaller than that of a normal distribution.

By searching iteratively for a set of thresholds corresponding to the three scores described above, we found a set of operating points that was continuous and devoid of holes: a simply connected region for which a hair bundle oscillated. Outside this region, the bundle was classified as quiescent. This classification scheme has the advantage that it is not based on the amplitude of a bundle's noisy displacement, which is difficult to determine for many operating points. To emphasize the fact that the amplitude of a bundle's noisy displacements was not used to classify operating points, quiescent operating points are illustrated with a color not found in the spectrum used to illustrate the amplitude of spontaneous oscillations.

The amplitude and frequency of spontaneous oscillations at each operating point corresponded to the main peak of the time trace's Fourier transform. To reduce spectral leakage owing to the short duration of each time trace, the trace was multiplied by a Hamming window after subtracting the mean displacement. The trace was then zero padded to improve the accuracy of determining the peak. Because the resulting Fourier transform contained spurious low-frequency and high-frequency peaks owing to the measurement system, we searched for the largest peak between a minimum of 0-2 Hz and a maximum of 100 Hz, well within the range of best frequencies expected for the bullfrog's sacculus *(34,35)*. We excluded the power-supply frequency of 60 Hz. To account for the change in the height of this peak owing to windowing, we rescaled the value by a factor determined by applying the procedure described above to a sinusoidal time trace with duration equal to that of the original trace.

To avoid including drift in the estimation of the root-mean-square (RMS) amplitude, we subtracted the mean from each time trace and found the local RMS magnitude for a moving window 1.5 times as large as the window used to remove the drift in the analysis of the displacement distribution. The drift was then removed from each of these time windows by subtracting a linear fit. The RMS magnitude of the entire time trace was defined to be the mean of the local RMS magnitudes.

**Sensitivity.** To evaluate a hair bundle's response to periodic stimuli, we calculated the bundle's phase-locked response $\langle \tilde{X}(\omega_s) \rangle$, in which $\tilde{X}(\omega_s)$ is the bundle's Fourier amplitude at the frequency of driving, $\omega_s$. Sensitivity was then defined as the modulus of the bundle's response function

$$\tilde{\chi}(\omega_s) = \frac{\langle \tilde{X}(\omega_s) \rangle}{\tilde{F}(\omega_s)} \,, \tag{12}$$

in which $\tilde{F}(\omega_s)$ is the Fourier amplitude of the stimulus force at the driving frequency.

**Phase difference with respect to stimulation.** For those instances in which a sinusoidal stimulus was provided, the phase difference between the sinusoidal component of motion commanded at the fiber's base $\Delta_C(t)$ and that of the hair bundle $X(t)$ was determined by the relation

$$\phi = \phi_\Delta - \phi_X = \tan^{-1}\frac{\Im(\tilde{\Delta}_C(\omega_S))}{\Re(\tilde{\Delta}_C(\omega_S))} - \tan^{-1}\frac{\Im(\tilde{X}(\omega_S))}{\Re(\tilde{X}(\omega_S))} \,, \tag{13}$$

in which $\tilde{X}(\omega_s)$ and $\tilde{\Delta}_C(\omega_s)$ are the Fourier transforms of the motions at the frequency of driving, $\omega_s$, and $\Re$ indicates the real and $\Im$ the imaginary part of a variable. A negative phase difference corresponds to a phase lead of the hair bundle with respect to the fiber and a positive difference corresponds to a phase lag. This convention was used to accord with previously published phase differences between a stimulus and the hair bundle's response *(10)*.

**Quality factor.** The response of a hair bundle often displayed resonance at some characteristic frequency $f_C$. To estimate the bundle's sharpness of tuning, we employed an operational definition of the quality factor

$$Q = \frac{f_C}{\Delta f} \,. \tag{14}$$

Here $\Delta f$ corresponds to the bandwidth at amplitude $A_C/\sqrt{2}$, in which $A_C$ is the amplitude of the bundle's response at the characteristic frequency. Larger values of $Q$ correspond to sharper resonance. In most cases this operational definition underestimated the quality of resonance.

**Vector strength.** To assess the degree of entrainment between the stimulus fiber and the hair bundle, we measured the vector strength between the two signals. To do so, we first calculated the Hilbert transform of each signal,

$$X_H(t) = F^{-1}\left[-i \cdot \text{sgn}(\omega) \cdot \tilde{X}(\omega)\right], \tag{15}$$

in which $F^{-1}$ is the inverse Fourier transform, *sgn* is the signum function, and $\tilde{X}$ is the Fourier transform of the bundle's motion. From the analytic signal $X_A(t) = X(t) + iX_H(t)$ we then calculated the instantaneous phase

$$\varphi(t) = \varphi_{\Delta_C}(t) - \varphi_X(t) = \tan^{-1}\frac{\Delta_{C,H}(t)}{\Delta_C(t)} - \tan^{-1}\frac{X_H(t)}{X(t)}, \tag{16}$$

in which $\Delta_{C,H}(t)$ is the Hilbert transform of the sinusoidal component of motion commanded at the fiber's base, $\Delta_C(t)$. The mean vector length, or vector strength from a signal of length $N$, is then given by

$$VS = \left|\frac{1}{N}\sum_{j=1}^{N} e^{i\varphi(t_j)}\right|, \tag{17}$$

in which $0 \leq VS \leq 1$. The vector strength equals one if two signals are identical and completely entrained and approaches zero as two signals become dissimilar in instantaneous phase. This parameter thus corresponds to the degree of entrainment between two signals. The angle of the mean vector is given by its argument.

**Hair-bundle modeling.** Simulations of a model of hair-bundle dynamics *(16)* were performed with Mathematica 9.0.0.0 and C. To mimic the stochastic effects observed experimentally, noise terms were added to the model to yield the equations

$$\dot{x} = a(x - f) - (x - f)^3 - Kx + F_c + F + \eta_x; \tag{18}$$

$$\tau \dot{f} = bx - f + \eta_f, \tag{19}$$

in which $x$ is the bundle's displacement, $f$ is the force owing to adaptation, $a$ is a negative stiffness owing to gating of the transduction channel, $\tau$ is the timescale of adaptation, $b$ is a compliance coupling bundle displacement to adaptation, $K$ is the sum of the bundle's load stiffness and pivot-spring stiffness, $F_c$ is the sum of the constant force intrinsic to the hair bundle and that owing to the load, and *F* is any time-dependent force applied to the bundle. All simulation results used the same values of *a, b,* and *τ* as our previous publication *(16)*. The additional white noise terms $\eta_x$ and $\eta_f$ were *δ*-correlated random variables drawn from Gaussian distributions; the standard deviations of these distributions are denoted $\sigma_x$ and $\sigma_f$. Equations 18 and 19 were integrated numerically by the Euler-Maruyama method. Because the model was designed to capture qualitative effects associated with active hair-bundle motility, the values of the variables and parameters have no quantitative meaning. To facilitate comparisons with the experimental results, however, we rescaled the displacements, frequencies, sensitivities, stiffnesses, and forces in figures.

The model qualitatively predicts the shape of the state diagram and the variation in oscillation amplitude and frequency within the region of spontaneous activity *(16)*. We created artificial state diagrams by simulating a stochastic version of the model for a set of load stiffnesses and constant forces to produce time traces similar to those recorded experimentally. To determine the parameter values for which the virtual bundles oscillated, we applied the same procedures used to analyze the experimental results. The procedure correctly identified the oscillatory regions when the noise was small (Supplementary Fig. 4A-B). When the noise was large, however, the model bundles switched between stable states for operating points in the low-stiffness region of the diagram. The bundles additionally exhibited noise-induced oscillations for points in an excitable region that was classified as quiescent when the noise was small (Supplementary Fig. 4C-D). These behaviors suggest that the oscillatory regions found from experimental data included behaviors other than true limit-cycle oscillations at low stiffness values. Taken together, the simulations accorded well with the experimental state diagrams and highlighted additional behaviors unmasked by biological noise (Fig. 2 and Supplementary Fig. 2).

**Statistics.** Paired Student's *t*-tests were used to determine the significance of amplitude responses, quality factors, and vector strengths; significance was defined as $p < 0.05$. Binned phase differences were evaluated with Rayleigh's test for non-uniformity of circular data.

A single number was used to describe the correlation between any two quantities for a given state diagram by finding Spearman's rank-correlation coefficient for all of the values of the two quantities in the oscillatory region of the diagram. Owing to the grid structure of the sampled state diagram, there were many duplicate values for the stiffness and constant force that in conjunction with correlations between multiple quantities in a state diagram limited the magnitude of any particular correlation coefficient. These ties were taken into account by a permutation test to determine the statistical significance of the correlation given the sampling structure and the data. The *p*-value for each coefficient was thus a better indication of the true significance of the correlation than the value of the coefficient itself.

## Acknowledgments

We thank Dr. P. Martin and the members of our research group for comments on the manuscript. JDS is supported by grants F30DC013468 and T32GM07739 from the National Institutes of Health. DÓM was a Research Associate and AJH is an Investigator of Howard Hughes Medical Institute.

## Authors' contributions

JDS, DÓM, BAF, and AJH designed the experiments. JDS, BAF, and MT performed the experiments. DÓM performed the simulations. JDS and DÓM analyzed the data. JDS, DÓM, and AJH wrote the manuscript. The authors declare no competing financial interests.

27. Frederickson-Hemsing, L., Strimbu, C.E., Roongthumskul, Y., Bozovic, D. (2012). Dynamics of Freely Oscillating and Coupled Hair Cell Bundles under Mechanical Deflection. *Biophys. J.* **102**, 1785-1792.

28. Kennedy, H.J., Crawford, A.C., Fettiplace, R. (2005). Force generation by mammalian hair bundles supports a role in cochlear amplification. *Nature* **433**, 880-883.

29. Fernandez, C., Goldberg, J.M. (1976). Physiology of Peripheral Neurons Innervating Otolith Organs of the Squirrel Monkey. I. Response to Static Tilts and to Long-Duration Centrifugal Force. *J. Neurophysiol.* **39**(5), 970-984.

30. Manley, G.A., Fay, R.R., Popper, A.N. (2007). *Active Processes and Otoacoustic Emissions in Hearing*. (Springer).

31. Choe, Y., Maganasco, M.O., Hudspeth, A.J. (1998). A model for amplification of hair-bundle motion by cyclical binding of Ca2+ to mechanoelectrical-transduction channels. *Proc. Natl. Acad. Sci. USA* **95**, 15321-15326.

32. Strogatz, S.H. (1994). *Nonlinear Dynamics and Chaos*. (Addison-Wesley).

33. Shlomovitz, R., Frederickson-Hemsing, L., Kao, A., Meenderink, S.W.F., Bruinsma, R., Bozovic, D. (2013). Low Frequency Entrainment of Oscillatory Bursts in Hair Cells. *Biophys. J.* **104**, 1661-1669.

34. Hironori, K., Lewis, E.R., Leverenz, E.L., Baird, R.A. (1982). Acute seismic sensitivity in the bullfrog ear. *Brain Res.* **250**, 168-172.

35. Yu, X., Lewis, E.R., Feld, D. (1991). Seismic and auditory tuning curves from bullfrog saccular and amphibian papillar axons. *J. Comp. Physiol. A*, 241-248.

36. Benser, M.E., Issa, N.P., Hudspeth, A.J. (1993). Hair-bundle stiffness dominates the elastic reactance to otolithic-membrane shear. *Hear. Res.* **68**, 243-252.

37. Robles, L., Ruggero, M.A. (2001). Mechanics of the Mammalian Cochlea. *Physiol. Rev.* **81**, 1305-1352.

38. Strimbu, C.E., Kao, A., Tokuda, J., Ramunno-Johsnon, D., Bozovic, D. (2010). State and evoked motility in coupled hair bundles of the bullfrog sacculus. *Hear. Res.* **265**, 38-45.

# Figure legends

**Figure 1. Measurement of experimental state diagrams with a mechanical-load clamp.** (A) A theoretical state diagram depicts the qualitative behavior of a hair bundle for different values of its load stiffness and constant force. These parameters determine whether a bundle will spontaneously oscillate, remain quiescent, or manifest bistable switching. A region of spontaneous oscillation is enclosed within a line of Hopf bifurcations (orange). Two Bautin points (squares) separate the supercritical portion of the line (thick) from the subcritical parts (thin). The hair bundle has one stable state and remains quiescent within the monostable region. In the bistable region bounded by the line of fold bifurcations (green), a bundle may switch between two stable states. The line of Hopf bifurcations intersects the line of fold bifurcations at two Bogdanov-Takens points (circles). (B) The amplitude of spontaneous oscillations is expected to increase as the load stiffness decreases (arrow). The amplitude's dependence on the constant force is more complex. Smaller amplitudes are denoted by darker shades of red. (C) The frequency of spontaneous oscillations is theorized to rise as the load stiffness grows. The frequency's dependence on the constant force load is more complex. Lower frequencies are denoted by darker shades of blue. (D) As shown in the magnified circular inset, the tip of a flexible glass stimulus fiber exerts a force on the kinociliary bulb at the top of a hair bundle while the position of the fiber's tip is projected onto a dual photodiode (dashed rectangles). Information from this displacement monitor is conveyed to a target computer, which compares the bundle's position with the displacement commanded by the host computer and provides feedback with gain to a piezoelectric actuator that displaces the fiber's base. The command signal and gain together define the stiffness and constant force confronting the hair bundle. Positive forces act toward the hair bundle's tall edge, to the right, and negative forces in the opposite direction.

**Figure 2. Experimental state diagrams of oscillatory hair bundles.** (A) The oscillations of a small hair bundle changed in character as the effective stiffness of the stimulus fiber increased; a few operating points elicited complex oscillations whose multimodal nature is captured by the experimental records. (B) An experimental state diagram shows the behavior of the same hair bundle for various combinations of load stiffness and constant force, encompassing most of the oval locus of spontaneous oscillation. The gray region corresponds to quiescent operating points. Within the ruddy locus of spontaneous oscillation, color intensity represents the root-mean-square (RMS) magnitude of oscillation. The colored circles in the associated panels mark the operating points in (A). (C) In another representation of the experimental state diagram the color intensity encodes the amplitude of oscillation. (D) A third depiction of the experimental state diagram for the same bundle portrays the frequency of oscillation for various combinations of load stiffness and constant force. (E) Experimental records show the motions of a medium-sized hair bundle. (F-H) Both the RMS magnitude (F) and the amplitude (G) of oscillation were smallest along the high-stiffness border of the oscillatory region. The colored circles in these panels represent the transect along which the records in (E) were obtained. (H) The oscillation frequency for the same hair bundle was greatest along the high-stiffness boundary of the oscillatory region. (I) A large hair bundle oscillated spontaneously for all combinations of constant force and load stiffness. (J-L) As the load stiffness increased, both the RMS magnitude (J) and amplitude (K) of spontaneous oscillation decreased. Colored circles correspond to the operating points whose experimental records are shown in (I). (L) Increasing the load stiffness evoked a corresponding increase in the frequency of oscillation. The actual stiffness and drag coefficient of the stimulus fiber were respectively $K_{SF} = 425~\mu\text{N}\cdot\text{m}^{-1}$ and $\xi_{SF} = 53~\text{nN}\cdot\text{s}\cdot\text{m}^{-1}$. Analysis parameters and statistics for each experimental state diagram can be found in Supplementary Table 1. Additional experimental state diagrams may be found in Supplementary Fig. 2.

**Figure 3. Hair-bundle responses to sinusoidal stimuli.** (A) The behavior of a medium-sized hair bundle in the absence of stimulation was first classified for different operating points. The bundle's response to sinusoidal stimulation was then analyzed as a function of stimulus frequency. Here the constant force was zero and the amplitude of the stimulus was 1.5 pN. The response peaked at 10 Hz for a stiffness of 300 µN·m$^{-1}$, with the amplitude and quality of the resonant peak decreasing as the load stiffness increased. When the bundle was exposed to 500 µM gentamicin, the frequency response lost its peak for a load stiffness of 300 µN·m$^{-1}$ (gray dashed line). (B) The phase of hair-bundle motion with respect to the corresponding stimuli is shown for the operating points defined in (A). A negative angle corresponds to the bundle's motion leading the stimulus. The dashed lines signify phase differences of ±90º. At a load stiffness of 300 µN·m$^{-1}$, the bundle's motion switched from a phase lead to a phase lag near the bundle's resonant frequency. This pattern disappeared for higher stiffnesses (orange and blue) and upon application of gentamicin (gray dashed line). (C) In a model of hair-bundle responsiveness with a intermediate level of noise (standard deviations of the noise terms $\sigma_x = 0.1$ and $\sigma_f = 0.1$), a bundle yielded responses similar to those in (A). The resonant peak was greatest for a stiffness of 339 near the boundary of the oscillatory region, which occurred for zero constant force and a load stiffness of 340 in the absence of noise. (D) In the same model, the phase of the bundle's motion with respect to that of the stimulus displayed a pattern similar to that for the oscillatory point in (B). The magnitudes of the maximum phase lead and phase lag peaked at a stiffness of 339. (E) The behavior of a small hair bundle in the absence of stimulation was first classified for different operating points. The sensitivity is portrayed as a function of stimulus force at 5 Hz, near the bundle's frequency of spontaneous oscillation. (F) The vector strength of the bundle's motion with respect to that of the stimulus is displayed for the same operating points as in (E). (G) Using the model described for panel (C), a virtual bundle's sensitivity is portrayed as a function of stimulus force for stimulus frequencies 10 % greater than the frequency of spontaneous oscillation. The pattern resembled that shown in (E). The light gray dashed line corresponds to a slope of -2/3. (h) The vector strength of the simulated bundle's

motion is plotted against stimulus force. The bundle was best entrained at a stiffness of 341 for a range of intermediate to large forces. The error bars for experiments represent standard errors of the means for four observations; those that are not shown are similar in magnitude to those that are included. For all experiments, the stiffness and damping coefficient of the stimulus fiber were respectively 425 $\mu N \cdot m^{-1}$ and 53 $nN \cdot s \cdot m^{-1}$. Additional examples can be found in Supplementary Figures 5 and 6. For the panels resulting from simulations, the stiffness, frequency, and force have been rescaled by a factor of 100 to facilitate comparison with the experimental data. Because the model was rescaled, no units are displayed for simulated results. All simulations used the same values of *a, b,* and *τ* as the original description of the theoretical model *(16)*. The error bars are calculated from three stochastic simulations. Values of the vector strength below 0.2 (shaded areas) correspond to regions with poor phase locking as quantified by the Rayleigh test.

**Figure 4. Hair-bundle entrainment by sinusoidal stimulation.** (A) The behavior of a medium hair bundle was first classified for different operating points in the absence of stimulation. The bundle's frequency response to stimuli of 0.5 pN in amplitude peaked at 10 Hz for all operating points. When the bundle was exposed to 500 μM gentamicin, its frequency response lacked a peak for load stiffnesses from 300 $\mu N \cdot m^{-1}$ to 800 $\mu N \cdot m^{-1}$ (dark and light dashed lines, respectively). (B) Quantified by the vector strength for each operating point in (A), the degree of entrainment peaked at a stiffness of 380 $\mu N \cdot m^{-1}$. (C) A second hair bundle oscillated at all three load stiffnesses of 100 $\mu N \cdot m^{-1}$ (red), 167 $\mu N \cdot m^{-1}$ (yellow) and 250 $\mu N \cdot m^{-1}$ (blue) (inset). The vector strengths for all operating points increased with stimulus force during stimulation at 5 Hz (solid lines). This stimulus frequency was selected to maximize the vector strength (Supplementary Fig. 7). When stimulated at 80 Hz, away from the frequency of spontaneous oscillation, the bundle was entrained poorly by the stimulus (dashed lines). (D) For a load stiffness of 250 $\mu N \cdot m^{-1}$ the bundle displayed a gradual decrease in the slope of the relation of vector strength to stimulus force as the frequency increased (5 Hz, 9 Hz, 21 Hz, and 80 Hz; dark

to light). (E) For a stimulus force of 6 pN, the same bundle achieved maximum entrainment at 5 Hz for a load stiffness of 250 µN·m$^{-1}$. Additional data are included in Supplementary Figures 7-9. (F-H) Heat maps display the vector strength as a function of stimulus force and stimulus frequency for stiffnesses of 100 µN·m$^{-1}$ (F), 167 µN·m$^{-1}$ (G), and 250 µN·m$^{-1}$ (H). The error bars represent standard errors of the mean for three observations; those not shown resembled in magnitude those that are included. The stiffness and damping coefficient of the stimulus fiber were respectively 425 µN·m$^{-1}$ and 53 nN·s·m$^{-1}$.

**Figure 5. Responses to force pulses.** (A) Movement of a stimulus fiber's base (black) subjected a large hair bundle under a load stiffness of 40 µN·m$^{-1}$ to a force pulse. For constant forces of -15 pN and -20 pN, the bundle's response (red) to the pulse (blue) displayed an increase in the rate of spontaneous oscillation. For a constant force of -25 pN, the bundle responded to a positive force pulse with a twitch and a negative force transient of 1.2 pN that decayed with a time constant of 5 ms (inset). (B) When a large hair bundle was subjected to a stiffness of 100 µN·m$^{-1}$ and a constant force of -66 pN, a positive force pulse elicited a response (red) smaller than the displacement of the fiber's base (black). The force applied by the fiber during the pulse (blue) was therefore positive. (C) When the constant force was increased to -100 pN, a positive force pulse (black) elicited a response (red) exceeding the displacement of the fiber's base; the force applied by the fiber (blue) was accordingly negative. The stiffness and damping coefficient of the stimulus fiber were respectively 105 µN·m$^{-1}$ and 71 nN·s·m$^{-1}$ (A) or 425 µN·m$^{-1}$ and 53 nN·s·m$^{-1}$ (B,C). The time traces have been downsampled by a factor of 100 for presentation purposes.

# Supplementary Information

for

# Control of a hair bundle's mechanosensory function by its mechanical load

Joshua D. Salvi*, Dáibhid Ó Maoiléidigh*, Brian A. Fabella, Melanie Tobin†, and A. J. Hudspeth

Howard Hughes Medical Institute and Laboratory of Sensory Neuroscience

The Rockefeller University, 1230 York Avenue, New York, New York, 10065, USA

**Supplementary figure legends**

**Supplementary Figure 1. Verification of the mechanical-load clamp.** (A) To verify that the clamp allowed for simultaneous control of the load stiffness and constant force, we used a stimulus fiber of stiffness $K_{SF} = 350$ µN·m$^{-1}$ and damping coefficient $\xi_{SF} = 164$ nN·s·m$^{-1}$ to deliver force steps to a vertically mounted glass fiber of stiffness $K_{HB} = 560$ µN·m$^{-1}$ that acted as a simulacrum of a hair bundle. For a given constant force and load stiffness, the steady-state position $X$ of the model bundle should behave as a Hookean material (Equation 11). The displacements of the test fiber (black circles) in response to forces delivered by a stimulus fiber are shown as a function of the added stiffness. For five levels of constant force, the application of a range of load stiffnesses yielded results demonstrating control of these parameters. Purple lines indicate fits to Equation 11. (B) To test the clamp's capability to hold a hair bundle at an operating point ($F_C = 0$ pN; $K_{EFF} = 150$ µN·m$^{-1}$ and 175 µN·m$^{-1}$) while stimulating a hair bundle sinusoidally, we delivered time-varying stimuli to another test fiber of stiffness $K_{SF} = 109$ µN·m$^{-1}$ and damping coefficient $\xi_{SF} = 133$ nN·s·m$^{-1}$. Stimulation at different frequencies yielded a response (black) that closely resembled that of a commanded signal (red).

(C) The ratio of the amplitude of the fiber's motion to the amplitude of the command signal deviates from the ideal by less than 2% at all frequencies at a gain of 0.3 (cyan) and 0.46 (purple). (D) The fiber's displacement lags the stimulus force to a small degree at all frequencies, increasing to 16º at 100 Hz. This dependence of phase on frequency is the same for a gain of 0.3 (cyan) and 0.46 (purple). The time traces have been downsampled by a factor of 100 for presentation purposes.

**Supplementary Figure 2. Additional hair-bundle experimental state diagrams.** We produced detailed experimental state diagrams for three hair bundles in addition to those analyzed in Figure 2. In the experimental state diagrams computed for two medium hair bundles (A-C, D-F) and one large hair bundle (G-I), all quiescent operating points are displayed in gray. For oscillatory operating points, the shades of red correspond to either the RMS magnitude or the amplitude of spontaneous oscillation. The shades of blue correspond to the frequency of spontaneous oscillations. For both medium bundles, we encountered a border along which oscillations were suppressed. Near the border, the peak frequencies were (C) 20 Hz and (F) 10 Hz. In the case of the large bundle, we were unable to suppress spontaneous oscillations. In many instances the amplitude was inversely correlated with the oscillation frequency (see Supplementary Table 1). The peak frequencies of spontaneous oscillations in these cells accorded with the range of characteristic frequencies for afferent nerve fibers from the bullfrog's sacculus.

**Supplementary Figure 3. State-diagram controls.** (A-B) As expected for an *in vitro* preparation, the activity of a hair bundle deteriorates gradually during protracted recording. If a hair cell were to exhibit significant changes in its state diagram over the course of an experiment, our conclusions would be compromised. To determine whether the results remained consistent over time, we computed a bundle's experimental state diagram at two times separated by 10 min. Contrary to the general practice, we did not exchange the artificial endolymph every 4-6 min.

The shades of red and blue correspond respectively to the amplitude and frequency of spontaneous oscillation. The data reveal little change in the bundle's state space over the course of the experiment, with correlation coefficients between the two diagrams of 0.92 ($p < 10^{-15}$) in amplitude and 0.86 ($p < 10^{-10}$) in frequency. The amplitude and frequency of the bundle's oscillation remained stable and the time traces did not yield obvious changes over the course of 10 min even without the regular change of artificial endolymph, verifying that the bundle's state diagram remains stable over the course of an experiment. (C-F) To better grasp the contribution of active hair-bundle motility on the bundle's state space, we computed the experimental state diagrams of two bundles bathed in 500 µM gentamicin, which blocks mechanotransduction channels and arrests spontaneous oscillations. (C) A control experimental state diagram was first measured for a large hair bundle. Oscillations had a maximal RMS magnitude of 16 nm and a mean of 8.7 nm. (D) During exposure to gentamicin the bundle became quiescent at all operating points. The amplitude map shows consistently small amplitudes, with a maximum RMS magnitude of 2.0 nm and a mean of 1.4 nm. The correlation between the experimental state diagrams before and after treatment was not significant, with coefficients of 0.21 ($p = 0.14$) in amplitude and 0.01 ($p = 0.95$) in frequency. (E) Another experimental state diagram was measured for a medium hair bundle, in which the bundle oscillated for half of the operating points. The greatest RMS magnitude was of 15.6 nm and the mean RMS magnitude was 4.5 nm. (F) Upon exposure to gentamicin the bundle became quiescent at all operating points, with a maximal RMS magnitude of 1.6 nm and a mean of 1.4 nm. As before, there was no significant correlation between the experimental state diagrams, with coefficients of 0.28 ($p = 0.13$) in amplitude and 0.15 ($p = 0.32$) in frequency. In all cases, the experimental state diagrams before and after gentamicin treatment showed no significant positive correlations. These controls verify that a hair bundle without active motility possesses no oscillatory operating points and that its experimental state diagram changes dramatically when active motility is abolished. The analysis and parameters for these experiments may be found in Supplementary Table 1.

**Supplementary Figure 4. Artificial state diagrams with noise.** (A) An artificial state diagram was generated in a model of hair-bundle mechanics with a low noise level: the standard deviations of the noise terms $\sigma_x = 0.001$ and $\sigma_f = 0.001$. The green lines correspond to a loop of Hopf bifurcations and a line of fold bifurcations. The gray operating points were classified as quiescent. Within the red region of spontaneous oscillation, color intensity corresponds to the amplitude of spontaneous oscillation. The smallest amplitudes were found near the high-stiffness border of the oscillatory region. (B) A second artificial state diagram under the same conditions depicts the oscillation frequency in blue. Near the edge of the region of spontaneous oscillation, frequencies reached their maximum. (C) Another artificial state diagram was generated with a high noise level: the standard deviations of the noise terms $\sigma_x = 1$ and $\sigma_f = 1$. As before, the amplitude of spontaneous oscillation is displayed in red and quiescent operating points are presented in gray. In this case, the region of spontaneous oscillation increased in size. (D) For the same noise level, an artificial state diagram presents the frequency of spontaneous oscillation in blue. As for the previous cell, the amplitude and frequency of spontaneous oscillation were inversely correlated, with the minimum amplitude and maximum frequency both occurring near the high-stiffness edge of the oscillatory region. The constant force, stiffness, displacement, and frequency have been rescaled by a factor of 100. Because the model incorporates rescaled parameters, no units for the amplitude, constant force, and load stiffness are displayed. All simulation results employed the same values of *a, b,* and *τ* as the original description of the theoretical model *(16)*. The analysis parameters and correlation statistics may be found in Supplementary Table 1.

**Supplementary Figure 5. Additional examples of active hair-bundle resonance.** (A,C) The behavior of two hair bundles was first classified for different operating points in the absence of stimulation. The hair bundle's amplitude of vibration in response to sinusoidal stimulation was then analyzed as a function of the stimulus frequency. The responses peaked at 40 Hz (A), 20 Hz (C). The largest and sharpest responses occurred for operating points near the boundary of the oscillatory region. (A) When the bundle was exposed to 500 μM gentamicin, the frequency response lost its peak for a load stiffness of 300 μN·m$^{-1}$ (gray dashed line). (B,D) The phase difference between the bundle's motion and that of the stimulus was calculated. A reversal from a phase lead to a phase lag occurred near the bundle's resonant frequency. This magnitude of the phase change associated with a reversal diminished upon increasing the load stiffness of the hair bundle, which moved its operating point farther from the edge of the oscillatory regions.

**Supplementary Figure 6. Additional examples of hair-bundle sensitivity.** (A) Stimuli of increasing magnitudes were delivered at frequencies near those of spontaneous oscillations to four hair bundles poised near the edges of their oscillatory regions. The bundle's load stiffness was decreased to coax its operating point farther into the oscillatory region. (B) For each operating point, entrainment to stimuli was quantified by vector strength. A value less than 0.2 (shaded area) corresponds to a region with poor phase locking as quantified by the Rayleigh test. In agreement with the plots of sensitivity, forces below 1 pN yielded a low vector strength and thus poor entrainment. As the force increased, the vector strength and thus the hair-bundle entrainment increased. The level at which the bundle's vector strength rose corresponded closely with a change in the slope of the force-sensitivity relation for all these cells.

**Supplementary Figure 7. Hair-bundle entrainment as a function of stimulus force across stimulus frequencies.** To further assess the degree of phase locking between a hair bundle and its stimulus, we held a hair bundle at load stiffnesses of (A) 100 μN·m$^{-1}$, (B) 167 μN·m$^{-1}$, and (C) 250 μN·m$^{-1}$, corresponding to operating points deep within its oscillatory region. Here the hair bundle exhibited relaxation oscillations of large amplitude and low frequency. We delivered stimuli of successively increasing forces at frequencies of 5, 9, 13, 17, 21, 24, 27, 30, 40, 60, and 80 Hz (dark to light). The degree of entrainment between the hair bundle's motion and that of the stimulus fiber is represented by the vector strength. A hair bundle achieved maximum phase locking for small forces at a stimulus frequency of 5 Hz.

**Supplementary Figure 8. Hair-bundle entrainment as a function of stimulus frequency for different stimulus forces.** A hair bundle oscillated at load stiffnesses of 100 µN·m$^{-1}$ (red), 167 µN·m$^{-1}$ (yellow), and 250 µN·m$^{-1}$ (blue). Sinusoidal forces were then delivered at amplitudes of (A) 1 pN, (B) 2 pN, (C) 3 pN, (D) 4 pN, (E) 6 pN, and (F) 9 pN. The degree of entrainment between the stimulus fiber's motion and that of the hair bundle is represented by the vector strength (Equation 16). For all stimulus forces, a hair bundle achieved maximum phase locking at a stimulus frequency of 5 Hz, near its frequency of spontaneous oscillation.

**Supplementary Figure 9. Hair-bundle entrainment as a function of stimulus frequency for different operating points.** A hair bundle oscillated at load stiffnesses of (A) 100 µN·m$^{-1}$, (B) 167 µN·m$^{-1}$, and (C) 250 µN·m$^{-1}$. Successively increasing stimulus forces were then delivered at frequencies of 5 Hz, 9 Hz, 13 Hz, 17 Hz, 21 Hz, 24 Hz, 27 Hz, 30 Hz, 40 Hz, 60 Hz, and 80 Hz. The degree of entrainment between the hair bundle's motion and that of the stimulus fiber is represented by the vector strength. For all stimulus forces, a bundle achieved maximum phase locking at a stimulus frequency of 5 Hz. The sharpness of the relation between vector strength and stimulus frequency peaked at 250 µN·m$^{-1}$. These data support the hypothesis that a hair bundle achieves optimal phase locking near the edge of the oscillatory region for small forcing and for small differences between frequency of stimulation and that of spontaneous oscillation.

**Supplementary Table 1. Summary of state-diagram analysis and statistics.** Statistics are displayed for each of the state diagrams in Figure 2 and Supplementary Figures 2, 3, and 4. The columns corresponding to Supplementary Figure 3 depict the statistics for one hair bundle over the course of 10 min (S3A-B) and two bundles prior to gentamicin treatment (S3C,E). For each hair bundle, its size classification and analysis parameters are presented (see Materials and Methods). Cells with dashed lines correspond to tests that were not required for that particular bundle. For each statistic, a threshold was placed both upon the statistic value presented and its corresponding *p*-value (*, $p < 0.01$; **, $p < 0.001$). Spearman's rank correlation between the bundle's RMS magnitude, amplitude, and frequency of spontaneous oscillations was calculated for each bundle with respect to both the load stiffness and constant force and between the amplitude and frequency of oscillation. Numbers in bold correspond to significant correlations ($p < 0.05$). In all cases a bundle's oscillatory amplitude decreased with increasing load stiffness. For half of the bundles the frequency of oscillation grew with increasing load stiffness. In half of the instances the amplitude and frequency of spontaneous oscillation were inversely correlated.